# Symmetry Origin and Microscopic Mechanism of Electrical Magnetochiral Anisotropy in Tellurium

Manuel Suárez-Rodríguez[1,2], Beatriz Martín-García[1,3], Francesco Calavalle[1], Stepan S. Tsirkin[3,4], Ivo Souza[3,4], Fernando de Juan[3,5], Albert Fert[2,5,6], Marco Gobbi[3,4,*], Luis E. Hueso[1,3,†], and Fèlix Casanova[1,3,‡]

[1]CIC nanoGUNE BRTA, 20018 Donostia-San Sebastián, Basque Country, Spain
[2]Department of Polymers and Advanced Materials: Physics, Chemistry and Technology UPV/EHU, 20018 Donostia-San Sebastián, Basque Country, Spain
[3]IKERBASQUE, Basque Foundation for Science, 48009 Bilbao, Basque Country, Spain
[4]Centro de Física de Materiales CSIC-UPV/EHU, 20018 Donostia-San Sebastián, Basque Country, Spain
[5]Donostia International Physics Center, 20018 Donostia-San Sebastián, Basque Country, Spain
[6]Laboratoire Albert Fert, CNRS, Thales, Université Paris-Saclay, 91767 Palaiseau, France

Correspondence to: *marco.gobbi@ehu.eus; †l.hueso@nanogune.eu; ‡f.casanova@nanogune.eu

*Abstract.-*Non-linear transport effects in response to external magnetic fields, i.e. electrical magnetochiral anisotropy (eMChA), have attracted much attention for their importance to study quantum and spin-related phenomena. Indeed, they have permitted the exploration of topological surface states and charge-to-spin conversion processes in low-symmetry systems. Nevertheless, despite the inherent correlation between the symmetry of the material under examination and its non-linear transport characteristics, there is a lack of experimental demonstration to delve into this relationship and to unveil their microscopic mechanisms. Here, we study eMChA in chiral elemental Tellurium (Te) along different crystallographic directions, establishing the connection between the different eMChA components and the crystal symmetry of Te. We observed different longitudinal eMChA components with collinear current and magnetic field, demonstrating experimentally the radial angular momentum texture of Te. We also measured a transverse non-linear resistance which, as the longitudinal counterpart, scales bilinearly with current and magnetic fields, illustrating that they are different manifestations of the same effect. Finally, we study the scaling law of the eMChA, evidencing that extrinsic scattering from dynamic sources is the dominant microscopic mechanism. These findings underscore the efficacy of symmetry-based investigations in understanding and predicting non-linear transport phenomena, with potential applications in spintronics and energy harvesting.

*Introduction.-*Low-symmetry materials have revolutionized the field of electronic transport. This disruption stems from the breaking of inversion symmetry within such systems, which permits non-linear transport effects where the voltage ($V$) scales quadratically with the current ($I$) [1]. Notably, this non-linear behavior can manifest both with and without the presence of external magnetic fields ($B$) [2].

On the one hand, non-linear transport effects in the absence of $B$ ($I \propto V^2$) have recently attracted much attention [3-6]. Indeed, they allowed for the study of novel quantum properties, such as the Berry curvature [7,8] or Berry connection polarizability [9,10], and have the potential to be exploited for energy harvesting through radiofrequency rectification [11-14]. Systematic studies have been performed, identifying both longitudinal (i.e., non-linear conductivity [15,16]) and transverse (i.e., non-linear Hall effect (NLHE) [3,4,17]) components, and its connection with the crystal symmetry of the tested material.

On the other hand, non-linear transport effects in the presence of $B$ ($I \propto V^2 B$) have great importance for spintronics as they can be employed for the investigation of spin-related effects [18]. In general, these phenomena are named electrical magnetochiral anisotropy (eMChA) [19,20]. However, as most reports are focused on the longitudinal manifestation, they are commonly known as unidirectional magnetoresistance (UMR) [18] or bilinear magnetoresistance (BMR) [21,22]. Although there are few reports on the transverse manifestation, i.e., non-linear planar Hall effect (NLPHE) [23,24], an experimental demonstration of the fundamental connection between the two effects and the crystal symmetry of the material is still absent. Furthermore, the microscopic mechanisms behind these effects remain unclear.

Chiral materials provide an exceptional platform for investigating non-linear transport effects due to their absence of both inversion and mirror symmetry [25-27]. In this context, a material with strong spin-orbit coupling such as chiral tellurium (Te) [28], which can be chemically synthesized [29] and patterned into desired shapes, emerges as the ideal candidate.

In this letter, we present an experimental study of eMChA in Te, encompassing both longitudinal and transverse measurements, along with an analysis of the crystallographic direction dependence. Te flakes are grown via a hydrothermal process and patterned into star-shaped structures, enabling precise measurements along different crystallographic directions. By taking into account the symmetry of trigonal Te, we derive equations that accurately describe the experimental non-linear transport in Te across all configurations. We observe a longitudinal eMChA, i.e. UMR or BMR, when $B$ and $I$ are aligned along the $z$-axis, consistently with previous reports [18]. Remarkably, we also observe a longitudinal eMChA when both $B$ and $I$ are along the $x$-axis, thus demonstrating the anisotropic radial angular momentum texture of Te by electrical means. Additionally, we detect a transverse eMChA, i.e., NLPHE, when $B$ is along the $z$-axis and $I$ is along the $x$-axis. We illustrate that this transverse non-linear resistance exhibits bilinear dependence on $B$ and $I$. Finally, by examining the dependence of the eMChA on the resistivity, we establish that the eMChA in Te is primarily governed by extrinsic mechanisms. Our findings underscore how the analysis of crystal symmetry and resistivity scaling facilitate the prediction of non-linear transport effects, providing insights into permitted components and microscopic mechanisms. Therefore, we aim to inspire similar analyses devoted to discovering novel systems suitable for spintronics and energy harvesting applications.

*Anisotropic transport.*-Trigonal elemental Te displays a chiral crystal structure, belonging to $P3_121$ (right-handed) or $P3_221$ (left-handed) space groups [Figs. 1(a),(b)]. We grow single crystalline Te flakes following a hydrothermal process [29], and we pattern them into a star-shaped device by e-beam lithography and reactive ion etching. Finally, we contact them with Pt electrodes allowing electrical measurements for different crystallographic directions (Supplementary Section 1 [30]). A harmonic current at frequency $\omega = 31$ Hz ($I^\omega$) is injected between two electrodes aligned in the same direction at a $\theta$-angle from the chiral $z$-axis. During this process, both the first- ($V^\omega$) and second- ($V^{2\omega}$) harmonic voltages are recorded. The measurements are conducted in both longitudinal ($V_\parallel$) and transverse ($V_\perp$) configurations using a rotating reference frame. Our setup also enables temperature modulation (2-300 K) and the application of $B$ up to 9 T in all directions [Fig. 1(c)].

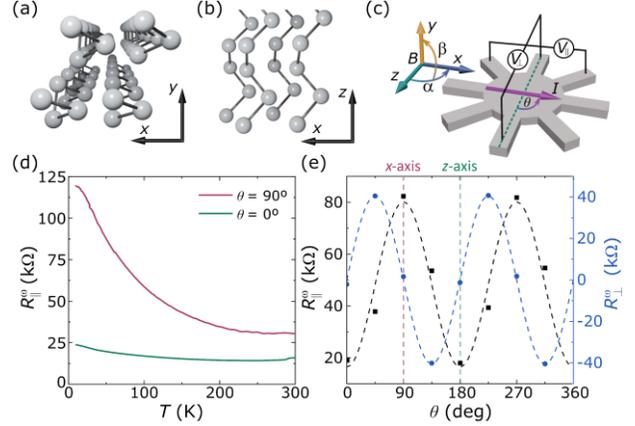

FIG 1. Anisotropic transport in Te. (a), (b) Crystal structure sketch of trigonal Te in the (a) $x-y$ and (b) $x-z$ plane. (c) Sketch of the star-shaped Te device. The relative orientation between $B$ and the device is indicated. (d) Temperature dependence of $R_\parallel^\omega$ when $I^\omega$ is injected along $x$- ($\theta = 90°$) and $z$-axis ($\theta = 0°$). (e) $R_\parallel^\omega$ and $R_\perp^\omega$ as a function of $\theta$ at 50 K. The dashed lines are fits to the equations introduced in the main text. The $R_\parallel^\omega$ and $R_\perp^\omega$ values are obtained from the linear fittings in Fig. S1 [30]. All measurements were performed at $B = 0$ T with $|I^\omega| = 1$ μA.

Figure 1(d) shows the temperature dependence of the first-harmonic longitudinal resistance $(R_\parallel^\omega \equiv \frac{V_\parallel^\omega}{I^\omega})$ along the $z$-axis ($\theta = 0°$) as well as the $x$-axis ($\theta = 90°$). We note that the resistance along the $x$-axis is much higher than along the $z$-axis, which is a direct consequence of the anisotropic crystal structure: the electronic transport is much more favorable along the covalently-bonded Te helices ($z$-axis) than between them ($x$-axis). To further study the anisotropic electronic transport in Te and to ensure a proper control of current directionality, we measure the first-harmonic longitudinal $(R_\parallel^\omega)$ and transverse resistance $(R_\perp^\omega \equiv \frac{V_\perp^\omega}{I^\omega})$ for different $\theta$ angles, between the current $I^\omega$ and the chiral $z$-axis [Fig. 1(e)]. On the one hand, $R_\parallel^\omega$ follows $R_\parallel^\omega(\theta) = (L_\parallel/A)[\rho_{zz}\cos^2(\theta) + \rho_{xx}\sin^2(\theta)]$, where $L_\parallel$ is the distance between longitudinal contacts, $A$ is the cross section of the contacts, and $\rho_{zz}$ and $\rho_{xx}$ are the resistivities along the z- and x-axis, respectively. We observe that $\rho_{xx} \gg \rho_{zz}$, in agreement with the temperature dependence study [Fig. 1(d)]. On the other hand, $R_\perp^\omega$ follows $R_\perp^\omega(\theta) = (L_\perp/A)(\rho_{xx} - \rho_{zz})\cos(\theta)\sin(\theta)$, where $L_\perp$ is the distance between transverse contacts. The excellent fittings of the experimental data to the expected $\theta$ dependence indicate a precise control of the current directionality in our devices, allowing for further crystallographic dependent transport experiments (see Supplementary Section 2 [30]).

*Anisotropic magnetotransport.-*Recently, the electronic transport properties of Te in response to $B$ have attracted much attention. Negative magnetoresistance [18,36] and planar Hall effect [37] have been reported, both being considered as possible signatures of Weyl physics [37]. However, it is worth noting that the Weyl points in trigonal Te are at ~0.4 and 0.5 eV below the top of the valence band [38], thus far away from where the electronic transport generally occurs [39]. In the following, we will demonstrate that, from crystal symmetry considerations, it is possible to analyze the different magnetoresistance and Hall components in trigonal Te.

The first-order electric field ($E_i^\omega$) in response to a current density ($j_j^\omega$) and a magnetic field ($B_{k,l}$) can be expressed through the material Magnetoresistance ($T_{ijkl}$) and Hall ($R_{ijk}$) tensors as $E_i^\omega = T_{ijkl}j_j^\omega B_k B_l + R_{ijk}j_j^\omega B_k$. We observe that the magnetoresistance and Hall contributions are even and odd, respectively, with respect to $B_{k,l}$. Therefore, it is possible to differentiate them experimentally. For Te, considering $E_i^\omega$ and $j_j^\omega$ in the $x-z$ plane and $B_{k,l}$ in all directions, the Magnetoresistance and Hall tensors have 9 and 2 non-zero elements, respectively [40]:

$$T_{ijkl}^{Te} = \begin{pmatrix} T_{xxxx} & T_{xxyy} & T_{xxzz} & T_{xxyz} & 0 & 0 \\ T_{zzxx} & T_{zzyy} & T_{zzzz} & 0 & 0 & 0 \\ 0 & 0 & 0 & 0 & T_{xzxz} & T_{xzxy} \end{pmatrix} \quad (1)$$

$$R_{ijk}^{Te} = \begin{pmatrix} 0 & 0 & 0 \\ 0 & R_{xzy} & 0 \\ 0 & -R_{xzy} & 0 \\ 0 & 0 & 0 \end{pmatrix} \quad (2)$$

Therefore, we can obtain the expressions of the longitudinal and transverse electric fields as a function of $\theta$, $\alpha$, and $\beta$ angles in terms of $T_{ijkl}$ and $R_{ijk}$ (see Supplementary Section 2 [30]). To disentangle the different components, we have measured the Te magnetoresistance $\left(MR \equiv \frac{R_\parallel^\omega(B=9T) - R_\parallel^\omega(B=0T)}{R_\parallel^\omega(B=0T)}\right)$ for $\theta = 0°$ [Fig. 2(a)] and $\theta = 90°$ [Fig. 2(b)], and also $R_\perp^\omega$ for $\theta = 0°$ [Fig. 2(c)] and $\theta = 90°$ [Fig. 2(d)], by rotating $B = 9$ T in both $\alpha$- and $\beta$- planes [Fig. 1(c)]. Figures 2(a) and 2(b) manifest that the equations obtained from our analysis based on Te crystal symmetry perfectly capture the experimental response, both for the $\alpha$-angle dependence (blue curves),

$$\text{MR} = (B^2/\rho_{zz})\cos^2(\theta)[T_{zzxx}\sin^2(\alpha) + T_{zzzz}\cos^2(\alpha)] + (B^2/\rho_{xx})\sin^2(\theta)[T_{xxxx}\sin^2(\alpha) + T_{xxzz}\cos^2(\alpha)], \quad (3)$$

and for the $\beta$-angle dependence (red curves),

$$\text{MR} = (B^2/\rho_{zz})\cos^2(\theta)[T_{zzyy}\sin^2(\beta) + T_{zzzz}\cos^2(\beta)] + (B^2/\rho_{xx})\sin^2(\theta)[T_{xxyz}\sin(\beta)\cos(\beta) + T_{xxyy}\sin^2(\beta) + T_{xxzz}\cos^2(\beta)], \quad (4)$$

(see Supplementary Section 2 [30]). Importantly, we recognized that the critical parameter for the MR is the direction of the magnetic field. The MR is negative when $B$ is along the z-axis ($\alpha = \beta = 0°$), independently on the direction of $I^\omega$ (see also Figs. S2(a),(b) [30]). Previous reports have studied the magnetotransport of Te only when $I^\omega$ is along the z-axis, and related its negative MR with the chiral anomaly [37]. However, the chiral anomaly can only apply for $B \parallel I^\omega$ [41], and we also observe negative MR when $B \perp I^\omega$. Therefore, our observation suggests a different mechanism behind the negative MR in Te, such as Berry curvature or orbital moments [42-45]. A temperature dependence of the MR can be found in Figs. S2(c),(d).

Regarding the transverse measurements [Figs. 2(c),(d)], we observe that the $\beta$-angle dependence ($\alpha=0°$, red curves) is given by the Hall component:

$$R_\perp^\omega = (L_\parallel/A)R_{xzy}B\sin(\beta)[\sin^2(\theta) + \cos^2(\theta)]. \quad (5)$$

Indeed, when the magnetic field is out-of-plane ($B \parallel y$-axis) we detect the ordinary Hall effect in Te, whose sign and magnitude indicate that the electronic transport is dominated by holes and permits to obtain the carrier density $n_h \approx 6.5 \times 10^{17}$ cm$^{-3}$ [Figs. S2(e),(f)]. For the $\alpha$-angle dependence ($\beta=0°$, blue curves), we observe the so-called planar Hall effect:

$$R_\perp^\omega = -2(L_\perp/A)T_{xzxz}B^2\sin(\alpha)\cos(\alpha)[\sin^2(\theta) - \cos^2(\theta)]. \quad (6)$$

the planar Hall effect is even respect to $B$, thus not being a true Hall effect [46]. The planar Hall effect has been considered a signature of Weyl physics [37], but we demonstrate that it is directly allowed by the symmetry of Te and its associated MR tensor ($T_{xzxz}$ component). Therefore, as for the negative MR, its origin may be related to Berry curvature or orbital moments. More importantly, the equations obtained from the analysis of Te symmetry perfectly capture all the experimental observations (solid lines in Fig. 2), demonstrating to be a powerful method. From the fittings of the experimental data, we determined the values of the resistivity, magnetoresistance, and Hall tensors components (Table S1 [30]).

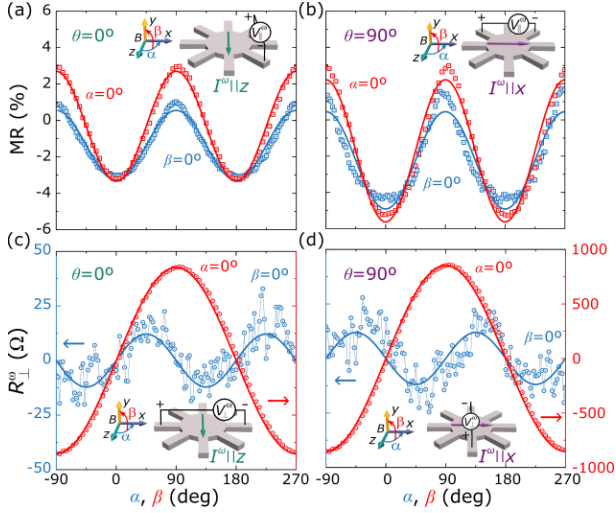

FIG 2. Anisotropic magnetotransport in Te. (a), (b) Te MR as a function of $\alpha$-angle ($\beta=0°$, blue curve) and $\beta$-angle ($\alpha=0°$, red curve) when $I^\omega$ is along (a) $z$-axis ($\theta = 0°$) and (b) $x$-axis ($\theta = 90°$). (c),(d) $R_\perp^\omega$ as a function of $\alpha$-angle ($\beta=0°$, blue curve) and $\beta$-angle ($\alpha=0°$, red curve) when $I^\omega$ is along (c) $z$-axis ($\theta = 0°$) and (d) $x$-axis ($\theta = 90°$). All measurements were performed at 50 K and $B$ = 9 T. The solid lines are fits to Eqs. (3)-(6).

*Electrical magnetochiral anisotropy.-*Non-linear transport effects in response to external magnetic fields, i.e. eMChA, have attracted much attention for their importance on spintronics. Indeed, eMChA has been reported in two-dimensional electron gases [22] and topological insulators [21], unveiling the helical spin texture of these systems. Recently, eMChA has also been reported in elemental Te [18]. The symmetry of chiral Te gives rise to an anisotropic handedness-dependent radial spin texture [28,47], which is compatible with the eMChA experiments. Therefore, the eMChA measurements can be exploited to determine the handedness of Te crystals [18]. These experiments have been limited to currents along the chiral $z$-axis. However, the Te chiral structure offers a rich family of eMChA components [20] that can be investigated with our star-shaped Te devices.

In general, the second-order current density ($j_i^{2\omega}$) in response to an electric ($E_{j,k}^\omega$) and magnetic ($B_l$) field can be expressed through the eMChA conductivity tensor ($\sigma_{ijkl}$) of the material as $j_i^{2\omega} = \sigma_{ijkl} E_j^\omega E_k^\omega B_l$.

For Te, considering $E_{j,k}^\omega$ and $j_i^{2\omega}$ in the $x-z$ plane and $B_l$ in all directions, the eMChA conductivity tensor has 8 independent non-zero elements [40]:

$$\sigma_{ijkl} = \begin{pmatrix} \sigma_{xxxx} & \sigma_{xzzx} & 0 \\ 0 & 0 & \sigma_{xxzy} \\ 0 & 0 & \sigma_{xxzz} \\ 0 & 0 & \sigma_{zxzx} \\ \sigma_{zxxy} & 0 & 0 \\ \sigma_{zxxz} & \sigma_{zzzz} & 0 \end{pmatrix} \quad (7)$$

As in the previous section, we can measure both the longitudinal and transverse response when $I^\omega$ is along $z$-axis ($\theta = 0°$) and $x$-axis ($\theta = 90°$). For such current directions, the longitudinal second-harmonic voltage ($V_\parallel^{2\omega}$) is given by (for more details, see Supplementary Section 3 [30]):

$$\theta = 0° \rightarrow V_\parallel^{2\omega} = -\frac{L_\parallel}{A^2}(I^\omega)^2 \rho_{zz}^3 \sigma_{zzzz} B\cos(\alpha)\cos(\beta) \quad (8)$$

$$\theta = 90° \rightarrow V_\parallel^{2\omega} = -\frac{L_\parallel}{A^2}(I^\omega)^2 \rho_{xx}^3 \sigma_{xxxx} B\sin(\alpha)\cos(\beta) \quad (9)$$

To explore the longitudinal components, we recorded $V_\parallel^{2\omega}$ as a function of the $\alpha$-angle at $B$ = 9 T and $\beta = 0°$ [Fig. 3(a)]. For $I^\omega \parallel z$-axis ($\theta = 0°$), the maximum signal is when $B$ and $I^\omega$ are collinear ($B \parallel I^\omega \parallel z$), as indicated by Eq. (8) and in agreement with previous reports [18]. Interestingly, for $I^\omega \parallel x$-axis ($\theta = 90°$), the maximum signal is also when $B$ and $I^\omega$ are collinear ($B \parallel I^\omega \parallel x$), as predicted by Eq. (9). Therefore, these results demonstrate the radial angular momentum texture of Te by purely electrical means. The eMChA is maximum when the external magnetic field is parallel to the current-induced angular momentum accumulation ($J_i$), which points along the $z$-axis when the current is along the $z$-axis ($J_z \parallel I^\omega \parallel z$), and points along the $x$-axis when the current is along the $x$-axis ($J_x \parallel I^\omega \parallel x$) [21]. The longitudinal eMChA is usually related to the spin texture of the system [21,22]. However, recent theoretical works suggest that the orbital contribution may be dominant [48]. Indeed, theoretical predictions indicate that in Te, the orbital ($L_i$) is stronger than the spin component ($S_i$) and, thus, dictates the eMChA response [20,49]. Therefore, as they have the same symmetry and we cannot distinguish between them in our experiments, we have decided to use the term "angular momentum texture" to account for both contributions ($J_i = L_i + S_i$). From the fittings, we obtain the values of $\sigma_{xxxx} = (-5.31 \pm 0.48) 10^{-6}$ AV$^{-2}$T and $\sigma_{zzzz} = (-1.283 \pm 0.048) 10^{-4}$ AV$^{-2}$T. The substantial difference between the magnitudes of $\sigma_{xxxx}$ and $\sigma_{zzzz}$ indicates that the angular momentum texture of Te is not merely radial but also anisotropic.

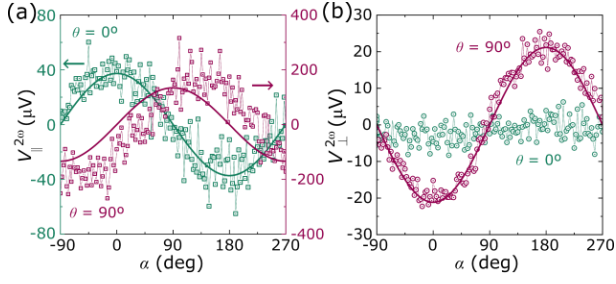

FIG 3. Crystallographic direction-dependence of eMChA in Te. (a) $V_{\parallel}^{2\omega}$ and (b) $V_{\perp}^{2\omega}$ when $I^{\omega}$ is along the $z$-axis ($\theta = 0°$, green curve) and the $x$-axis ($\theta = 90°$, magenta curve) as a function of $\alpha$-angle at 9 T and $\beta = 0°$. All measurements were performed at 50 K with $|I^{\omega}| = 1$ μA. The solid lines are fits of the experimental data to Eqs. (8)-(11).

The transverse eMChA, i.e. NLPHE, has been much less studied in literature [23,24]. Indeed, current direction-dependent studies, which identify the connections between the eMChA components and the crystal symmetry of the studied material, are still missing. Here, we employ our analysis based on Te symmetry, which determines that the transverse second-harmonic voltage ($V_{\perp}^{2\omega}$) is given by (see Supplementary Section 3 [30]):

$$\theta = 0° \rightarrow V_{\perp}^{2\omega} = -\frac{L_{\perp}}{A^2}(I^{\omega})^2 \rho_{xx}\rho_{zz}^2 \sigma_{xzzz} B\sin(\alpha)\cos(\beta) \quad (10)$$

$$\theta = 90° \rightarrow V_{\perp}^{2\omega} = -\frac{L_{\perp}}{A^2}(I^{\omega})^2 \rho_{xx}^2\rho_{zz} B[\sigma_{zxxz}\cos(\alpha)\cos(\beta) + \sigma_{zxxy}\sin(\beta)] \quad (11)$$

Hence, to explore the transverse components, we record $V_{\perp}^{2\omega}$ as a function of the $\alpha$-angle at $B$ = 9 T [Fig. 3(b)]. We note that, in the $\alpha$-angle dependence at $\beta = 0°$, the component $\sigma_{zxxy}$ does not contribute [see Eq. (11)]. Remarkably, we observe a clear eMChA signal when $I^{\omega} \parallel x$ (magenta curve, $\theta = 90°$) but not when $I^{\omega} \parallel z$ (green curve, $\theta = 0°$). Assuming $\sigma_{zxxz}$ and $\sigma_{xzzx}$ to be of the same order of magnitude, $V_{\perp}^{2\omega}$ is expected to be larger for $I^{\omega} \parallel x$, because $\rho_{xx} \gg \rho_{zz}$ [see equations (10) and (11)]. Moreover, $V_{\perp}^{2\omega}$ in response to $I^{\omega}$ and $B$ has also a contribution coming from a combination of non-linear conductivity and ordinary Hall. For $\beta = 0°$, this extra contribution has no impact when $I^{\omega} \parallel x$ but does have one when $I^{\omega} \parallel z$, and, therefore, it may cancel out the eMChA component (see Supplementary Section 4 [30]). From the fitting to Eq. (11), we quantify the value of $\sigma_{zxxz} = (2.603 \pm 0.039)\, 10^{-6}$ AV$^{-2}$T. More importantly, for both longitudinal and transverse measurements, our symmetry analysis perfectly capture the experimental response (solid lines in Fig. 3), unveiling the relationships between voltage, current and magnetic field directions for which the eMChA in Te is allowed.

We further explore the transverse eMChA, by studying its current-, field-, and temperature-dependence. The longitudinal eMChA, when defined as a resistance ($R_{\parallel}^{2\omega} = V_{\parallel}^{2\omega}/I^{\omega}$), is commonly known as BMR because it depends linearly on $I^{\omega}$ and $B$ [Eqs. (8)-(9)]. However, as dictated by Eqs. (10)-(11), the transverse non-linear resistance ($R_{\perp}^{2\omega} = V_{\perp}^{2\omega}/I^{\omega}$) is also expected to depend linearly on both. To prove this, we record $V_{\perp}^{2\omega}$ for different $I^{\omega}$ at $B$ = 9 T [Fig. 4(a)] and for different $B$ at $|I^{\omega}| = 1$ μA [Fig. 4(b)], as a function of $\beta$-angle at $\alpha = 0°$ and $\theta = 90°$. We note that, to fully demonstrate the accuracy of our symmetry analysis, we have studied $V_{\perp}^{2\omega}$ as a function of $\alpha$-angle in Fig. 3 and as a function of $\beta$-angle in Fig. 4. As observed, the experimental behavior of $V_{\perp}^{2\omega}$ is perfectly captured by Eq. (11), and $R_{\perp}^{2\omega}$ depends bilinearly on current and magnetic field [insets in Fig. 4(a),(b)]. The same bilinear dependence is obtained with the results as a function of $\alpha$-angle (Fig. S3 [30]), providing further evidence of the accuracy of our symmetry analysis. Eq. (11) includes a term $\sigma_{zxxy}\sin(\beta)$, but experimentally no $\sin(\beta)$ dependence is observed. The reason may be that $\sigma_{zxxy}$ is negligible respect to $\sigma_{zxxz}$ or its impact on $V_{\perp}^{2\omega}$ could be cancelled out by the contribution of the non-linear conductivity combined with the ordinary Hall effect, which is also allowed for $\theta = 90°$ with the same $\sin(\beta)$ dependence (see Supplementary Section 4 [30]).

Finally, we studied the microscopic mechanism of the eMChA [2]. For that purpose, we adapted the analysis developed for time-odd nonlinear transport in magnetic systems [50]. Indeed, the external magnetic field in eMChA plays a similar role to the internal magnetic vectors in time-odd nonlinear transport within magnetic systems. However, in the eMChA analysis, the orbital magnetic field contributions must also be included [20] (for more details, see Supplementary Section 5 [30]). The procedure relies on examining the scaling law between the output voltage, in this case $V_{\perp}^{2\omega}$, and the resistivity of the material, $\rho_{xx}$:

$$\frac{V_{\perp}^{2\omega}}{(I^{\omega})^2} = \gamma \rho_{xx}^{-1} + \delta + \xi \rho_{xx} + \zeta \rho_{xx}^2 + \eta \rho_{xx}^3 \quad (12)$$

where $\eta$ is a resistivity-independent parameter, while $\gamma$, $\delta$, $\xi$ and $\zeta$ only depend on the residual resistivity of the material. The intrinsic contributions of the eMChA are included in $\xi$ and $\eta$ parameters (see Supplementary Section 5 [30]). To explore the eMChA scaling law in our Te device, we modulate the resistivity by varying $T$ [Fig. 1(d)]. Within this $T$ range, we recorded $V_\perp^{2\omega}$ for $I^\omega$ along $x$-axis ($\theta = 90°$) as a function of the $\beta$-angle at $B$ = 9 T and $\alpha = 0°$ [Fig. 4(c)]. Hence, we can now represent $V_\perp^{2\omega}$ as a function of $T$ [inset in Fig. 4(c)] and, thus, as a function of $\rho_{xx}$ [Fig. 4(d)]. By fitting the experimental data to Eq. (12), we discern that $\zeta$ term dominates, demonstrating that an extrinsic mechanism governs the eMChA in Te (see Supplementary Section 5 [30]). We performed the same analysis for the longitudinal eMChA with $I^\omega$ along the $z$-axis ($\theta = 0°$), obtaining the same conclusions (Fig. S4, Table S2 [30]). Therefore, we have successfully identified the dominant microscopic mechanism in our Te devices. This methodology paves the way for similar analyses to uncover the microscopic mechanisms behind eMChA in a wide range of non-centrosymmetric systems.

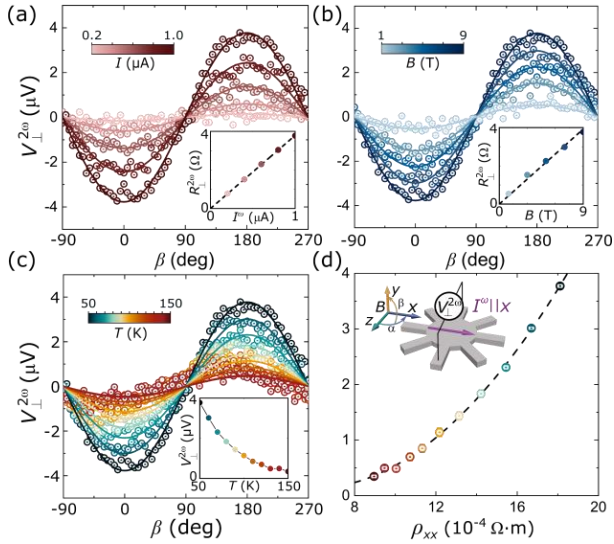

FIG 4. Current-, field- and temperature-dependence of the eMChA in Te. (a), (b), (c) $V_\perp^{2\omega}$ when $I^\omega$ is along the $x$-axis ($\theta = 90°$) as a function of $\beta$-angle for (a) different $I^\omega$ at $B$ = 9 T and $T$ = 50 K, (b) different $B$ at $T$ = 50 K and $|I^\omega|$ = 1 μA, and (c) different $T$ at $B$ = 9 T and $|I^\omega|$ = 1 μA. (d) $V_\perp^{2\omega}$ when $I^\omega$ is along the $x$-axis ($\theta = 90°$), obtained from the fittings in panel (c) to Eq. (11), as a function of $\rho_{xx}$. The dashed line is a fit to Eq. (12). Insets: (a) $I^\omega$ and (b) $B$ dependence of $R_\perp^{2\omega} = V_\perp^{2\omega}/I^\omega$, (c) $T$ dependence of $V_\perp^{2\omega}$. The values have been obtained from the fittings in panels (a), (b), and (c) to Eq. (11). (d) Sketch of the measurement configurations.

**Acknowledgements**

This work is supported by the Spanish MICIU/AEI/10.13039/501100011033 and by ERDF/EU (Projects No. PID2021-122511OB-I00, No. PID2021-128760NB-I00, No. PID2021-129035NB-I00, and "Maria de Maeztu" Units of Excellent Programme No. CEX2020-001038-M). It is also supported by MICIU and by the European Union NextGenerationEU/PRTR-C17.I1, as well as by IKUR Strategy under the collaboration agreement between Donostia International Physics Center and CIC nanoGUNE on behalf of the Department of Education of the Basque Government. M. S.-R. acknowledges support from La Caixa Foundation (No. 100010434) with code LCF/BQ/DR21/11880030. B. M.-G. and M. G. thank support from the "Ramón y Cajal" Programme by the Spanish MICIU/AEI and European Union NextGenerationEU/PRTR (Grants No. RYC2021-034836-I and No. RYC2021-031705-I). A. F. acknowledges the support of the UNIVERSIDAD DEL PAIS VASCO as distinguished researcher.

Supplemental material for:

# Symmetry Origin and Microscopic Mechanism of Electrical Magnetochiral Anisotropy in Tellurium

Manuel Suárez-Rodríguez, Beatriz Martín-García, Franceso Calavalle, Stepan S. Tsirkin, Ivo Souza, Fernando De Juan, Albert Fert, Marco Gobbi, Luis E. Hueso, and Fèlix Casanova.

## Contents

# Supplementary Section 1. Methods

*I. Chemical Synthesis of Te flakes.*-As in our previous paper [16], we synthesized the Te flakes by chemically reducing sodium tellurite ($Na_2TeO_3$, 99%) in presence of hydrazine ($N_2H_4$) in a basic aqueous medium at high temperature. We dissolved at room temperature $Na_2TeO_3$ (104 mg) and polyvinylpyrrolidone (average $M_w$ 29000-PVP 29, 92.1 mg) in 33 mL of type I water by magnetic stirring. Then, $NH_4OH$ solution (3.65 mL, 25% w in water) and hydrazine hydrate (1.94 mL, 80%, w/w%) were added while stirring. The mixture was transferred to an autoclave that was sealed and heated at 180 °C for 23 h. We purified the resulting material by successive centrifuge-assisted precipitation (10000 rpm-1h, Avanti J-26 XPI centrifuge) and redispersion with type I water. Finally, the Te flakes were redispersed in a dimethylformamide (DMF, $\geqslant$ 99.8 %) and $CHCl_3$ ($\geqslant$ 99.8 %) mixture (1.3:1 v/v) to be used in the sample fabrication. All the reagents were purchased from Sigma Aldrich and used as received without any further purification.

*II. Sample fabrication.*-Te flakes were redispersed in a dimethylformamide/$CHCl_3$ mixture (1.3:1 v/v) to be used for the drop-casting of solution droplets at the type I-water/air interface in a homemade Langmuir trough. After the evaporation of the solvent, Te flakes floating on the water surface were picked up (Langmuir–Schaefer technique) with Si/$SiO_2$ substrates (Si doped n+, 5 x 5 mm, 300 nm thermal oxide). Isolated flakes with suitable dimensions were selected with an optical microscope, without knowing a priori the handedness of the flakes. The flake shape and the contacts were defined through electron-beam lithography performed on a poly(methylmethacrylate)-A4/poly(methylmethacrylate)-A2 double layer. The patterning of the flakes was performed by reactive ion etching (Ar gas, 50 W RF power, 5 x 2 min), and Pt was deposited by sputtering for the electrical contacts. Te has a very anisotropic electrical resistivity and, therefore, the patterning is needed to properly control the current directionality [Fig. S1(a)].

*III. Electrical measurements.*-The devices were wire-bonded to a sample holder and installed in a physical property measurement system (by Quantum Design) for transport measurements with a temperature range of 2-400 K and maximum magnetic field of 9 T. We used a Keithley 6221 to apply a.c. current with a frequency of 31 Hz (ranging from 200 nA to 1 $\mu A$) and the longitudinal and transverse voltage drops were measured at the fundamental and the second-harmonic frequencies with a dual channel NF LI5660 lock-in amplifier.

In order to obtain reliable MR, Hall, and eMChA parameters from the angle-dependent measurements, it is crucial to remove any undesired contribution from unavoidable sample tilting (at 9 T even a tiny angle can give a significant contribution). In the first-harmonic transverse experiments, a tilt between the sample and the holder entangles the contributions from ordinary Hall and planar Hall effects. Therefore, to address this issue: (1) in the $\alpha$-dependence experiments, we measure at positive and negative magnetic fields and calculate the average signal to remove the contribution of the ordinary Hall effect caused by the tilting (which gives a finite, but not constant, $\beta$ angle) and obtain the clean planar Hall effect signal, and (2) in the $\beta$-dependence experiments, we also measure at positive and negative magnetic fields, but in this case, we calculate the half-difference signal to remove the contribution of the planar Hall effect caused by the tilting (which gives a finite, but not constant, $\alpha$ angle) and obtain the ordinary Hall effect signal. By this process, the two effects are disentangled. In the second-harmonic experiments, Te shows a non-linear conductivity at zero-order in $B$. Hence, to remove the contribution of the non-linear conductivity and to focus on the study of eMChA, we measure at positive and negative magnetic fields and calculate the half-difference signal. This procedure removes the non-linear conductivity contribution, which is at zero-order in $B$, and leaves the eMChA contribution, which is at first-order in $B$, unaffected.

## Supplementary Section 2. First-order electronic transport of Tellurium

**Electrical resistance**

The first-order electric field $E_i^\omega$ in response to a current $j_j^\omega$ in the absence of external magnetic fields can be expressed through the material's resistivity tensor $\rho_{ij}$ as $E_i^\omega = \rho_{ij} j_j^\omega$. For Te with $P3_121$ or $P3_221$ space group symmetry, the resistivty tensor (considering the electric field and the current in the $x-z$ plane) has 2 non-zero elements:

$$\rho_{ij}^{Te} = \begin{pmatrix} \rho_{xx} & 0 \\ 0 & \rho_{zz} \end{pmatrix} \tag{S1}$$

Therefore, the electric field along $x-$ and $z-$axis in response to a current $j_j^\omega$ can be written as:

$$E_x^\omega = \rho_{xx} j_x^\omega \tag{S2}$$

$$E_z^\omega = \rho_{zz} j_z^\omega \tag{S3}$$

For a rotating current in the plane $x-z$, $j_i^\omega = j^\omega(sin(\theta), 0, cos(\theta))$:

$$E_x^\omega = \rho_{xx} j^\omega sin(\theta) \tag{S4}$$

$$E_z^\omega = \rho_{zz} j^\omega cos(\theta) \tag{S5}$$

Separating in longitudinal and transverse components:

$$E_\parallel^\omega = j^\omega [\rho_{xx} sin^2(\theta) + \rho_{zz} cos^2(\theta)] \tag{S6}$$

$$E_\perp^\omega = j^\omega [\rho_{xx} - \rho_{zz}] sin(\theta) cos(\theta) \tag{S7}$$

Now, using: $E_\parallel^\omega = V_\parallel^\omega / L_\parallel$, $E_\perp^\omega = V_\perp^\omega / L_\perp$ and $j^\omega = I^\omega / A$, where $L_\parallel$ and $L_\perp$ corresponds to the distance between parallel and transverse contacts and $A$ to the cross section, we can rewrite Eqs. (S6) and (S7) in terms of the parameters that we directly measure in our experiments:

$$V_\parallel^\omega = \frac{L_\parallel}{A} I^\omega [\rho_{xx} sin^2(\theta) + \rho_{zz} cos^2(\theta)] \tag{S8}$$

$$V_\perp^\omega = \frac{L_\perp}{A} I^\omega [\rho_{xx} - \rho_{zz}] sin(\theta) cos(\theta) \tag{S9}$$

**Magnetoresitance**

The first-order electric field $E_i^\omega$ in response to a current $j_j^\omega$ and a magnetic field $B_{k,l}$ can be expressed through the material's magnetoresistance tensor $T_{ijkl}$ as $E_i^\omega = T_{ijkl} j_j^\omega B_k B_l$. For Te with $P3_121$ or $P3_221$ space group symmetry, the magnetoresistance tensor (considering the electric field and the current in the $x-z$ plane and the magnetic field in all directions) has 9 non-zero elements [40]:

$$T_{ijkl}^{Te} = \begin{pmatrix} T_{xxxx} & T_{xxyy} & T_{xxzz} & T_{xxyz} & 0 & 0 \\ T_{zzxx} & T_{zzyy} & T_{zzzz} & 0 & 0 & 0 \\ 0 & 0 & 0 & 0 & T_{xzxz} & T_{xzxy} \end{pmatrix} \tag{S10}$$

Therefore, the electric field along $x-$ and $z-$axis in response to a current $j_j^\omega$ and a magnetic field $B_{k,l}$ can we written as:

$$E_x^\omega = T_{xxxx} j_x^\omega B_x^2 + T_{xxyy} j_x^\omega B_y^2 + T_{xxzz} j_x^\omega B_z^2 + T_{xxyz} j_x^\omega B_y B_z + T_{xzxz} j_z^\omega B_x B_z + T_{xzxy} j_z^\omega B_x B_y \tag{S11}$$

$$E_z^\omega = T_{zzxx} j_z^\omega B_x^2 + T_{zzyy} j_z^\omega B_y^2 + T_{zzzz} j_z^\omega B_z^2 + T_{xzxz} j_x^\omega B_x B_z + T_{xzxy} j_x^\omega B_x B_y \tag{S12}$$

Thus, for a rotating magnetic field in all directions, $B_i = B(sin(\alpha)cos(\beta), sin(\beta), cos(\alpha)cos(\beta))$:

$$E_x^\omega = T_{xxxx}j_x^\omega B^2 sin^2(\alpha)cos^2(\beta) + T_{xxyy}j_x^\omega B^2 sin^2(\beta) + T_{xxzz}j_x^\omega B^2 cos^2(\alpha)cos^2(\beta) +$$
$$T_{xxyz}j_x^\omega B^2 cos(\alpha)sin(\beta)cos(\beta) + T_{xzxz}j_z^\omega B^2 sin(\alpha)cos(\alpha)cos^2(\beta) + T_{xzxy}j_z^\omega B^2 sin(\alpha)sin(\beta)cos(\beta) \tag{S13}$$

$$E_z^\omega = T_{zzxx}j_z^\omega B^2 sin^2(\alpha)cos^2(\beta) + T_{zzyy}j_z^\omega B^2 sin^2(\beta) + T_{zzzz}j_z^\omega B^2 cos^2(\alpha)cos^2(\beta) +$$
$$T_{xzxz}j_x^\omega B^2 sin(\alpha)cos(\alpha)cos^2(\beta) + T_{xzxy}j_x^\omega B^2 sin(\alpha)sin(\beta)cos(\beta) \tag{S14}$$

If now we consider a rotating current in the plane $x-z$, $j_i^\omega = j^\omega(sin(\theta), 0, cos(\theta))$, we obtain:

$$E_x^\omega = 2j^\omega cos(\theta)B^2[T_{xzxz}sin(\alpha)cos(\alpha)cos^2(\beta) + T_{xzxy}sin(\alpha)sin(\beta)cos(\beta)]$$
$$+j^\omega sin(\theta)B^2[T_{xxxx}sin^2(\alpha)cos^2(\beta) + T_{xxyy}sin^2(\beta) + T_{xxzz}cos^2(\alpha)cos^2(\beta) + T_{xxyz}cos(\alpha)sin(\beta)cos(\beta)] \tag{S15}$$

$$E_z^\omega = 2j^\omega sin(\theta)B^2[T_{xzxz}sin(\alpha)cos(\alpha)cos^2(\beta) + T_{xzxy}sin(\alpha)sin(\beta)cos(\beta)]$$
$$+j^\omega cos(\theta)B^2[T_{zzxx}sin^2(\alpha)cos^2(\beta) + T_{zzyy}sin^2(\beta) + T_{zzzz}cos^2(\alpha)cos^2(\beta)] \tag{S16}$$

Finally, separating in longitudinal and transverse components:

$$E_\parallel^\omega = 2j^\omega sin(\theta)cos(\theta)B^2[T_{xzxz}sin(\alpha)cos(\alpha)cos^2(\beta) + T_{xzxy}sin(\alpha)sin(\beta)cos(\beta)] +$$
$$j^\omega cos^2(\theta)B^2[T_{zzxx}sin^2(\alpha)cos^2(\beta) + T_{zzyy}sin^2(\beta) + T_{zzzz}cos^2(\alpha)cos^2(\beta)] + \tag{S17}$$
$$j^\omega sin^2(\theta)B^2[T_{xxxx}sin^2(\alpha)cos^2(\beta) + T_{xxyy}sin^2(\beta) + T_{xxzz}cos^2(\alpha)cos^2(\beta) + T_{xxyz}cos(\alpha)sin(\beta)cos(\beta)]$$

$$E_\perp^\omega = 2j^\omega[cos^2(\theta) - sin^2(\theta)]B^2[T_{xzxz}sin(\alpha)cos(\alpha)cos^2(\beta) + T_{xzxy}sin(\alpha)sin(\beta)cos(\beta)] +$$
$$j^\omega sin(\theta)cos(\theta)B^2[T_{xxxx}sin^2(\alpha)cos^2(\beta) + T_{xxyy}sin^2(\beta) + T_{xxzz}cos^2(\alpha)cos^2(\beta) + \tag{S18}$$
$$T_{xxyz}cos(\alpha)sin(\beta)cos(\beta) - T_{zzxx}sin^2(\alpha)cos^2(\beta) - T_{zzyy}sin^2(\beta) - T_{zzzz}cos^2(\alpha)cos^2(\beta)]$$

**Hall contribution**

Now, we will include also the Hall contribution. The first-order Hall response to a current $j_j^\omega$ and a magnetic field $B_k^\omega$ can be expressed through the material's Hall tensor $R_{ijk}$ as $E_i^\omega = R_{ijk}j_j^\omega B_k$. For Te with $P3_121$ or $P3_221$ space group symmetry, the Hall tensor (considering the electric field and the current in the $x-z$ plane and the magnetic field in all directions) has 2 non-zero elements [40]:

$$R_{ijk}^{Te} = \begin{pmatrix} 0 & 0 & 0 \\ 0 & R_{xzy} & 0 \\ 0 & -R_{xzy} & 0 \\ 0 & 0 & 0 \end{pmatrix} \tag{S19}$$

Therefore, the Hall contribution of the electric field along $x-$ and $z-$axis can we written as:

$$E_x = R_{xzy}j_z B_y \tag{S20}$$
$$E_z = -R_{xzy}j_x B_y \tag{S21}$$

Thus, for a rotating magnetic field in all directions $B_i = B(sin(\alpha)cos(\beta), sin(\beta), cos(\alpha)cos(\beta))$:

$$E_x^\omega = R_{xzy}j_z^\omega B sin(\beta) \tag{S22}$$
$$E_z^\omega = -R_{xzy}j_x^\omega B sin(\beta) \tag{S23}$$

If now we consider a rotating current in the plane $x - z$, $j_i^\omega = j^\omega(sin(\theta), 0, cos(\theta))$:

$$E_x^\omega = R_{xzy}j^\omega cos(\theta) Bsin(\beta) \tag{S24}$$
$$E_z^\omega = -R_{xzy}j^\omega sin(\theta) Bsin(\beta) \tag{S25}$$

Finally, separating in longitudinal and transverse components:

$$E_\parallel^\omega = 0 \tag{S26}$$
$$E_\perp^\omega = R_{xzy}j^\omega[sin^2(\theta) + cos^2(\theta)]Bsin(\beta) \tag{S27}$$

As expected, the Hall contribution is purely transverse.

**All first-order contributions**

If we take into account all first-order contributions: electrical resistance, magnetoresistance and Hall, we obtain:

$$\begin{aligned}
E_\parallel^\omega = j^\omega[\rho_{xx}sin^2(\theta) + \rho_{zz}cos^2(\theta)]+ \\
2j^\omega sin(\theta)cos(\theta)[T_{xzxz}B^2 sin(\alpha)cos(\alpha)cos^2(\beta) + T_{xzxy}sin(\alpha)sin(\beta)cos(\beta)]+ \\
j^\omega cos^2(\theta)B^2[T_{zzxx}sin^2(\alpha)cos^2(\beta) + T_{zzyy}sin^2(\beta) + T_{zzzz}cos^2(\alpha)cos^2(\beta)]+ \\
j^\omega sin^2(\theta)B^2[T_{xxxx}sin^2(\alpha)cos^2(\beta) + T_{xxyy}sin^2(\beta) + T_{xxzz}cos^2(\alpha)cos^2(\beta) + T_{xxyz}cos(\alpha)sin(\beta)cos(\beta)]
\end{aligned} \tag{S28}$$

$$\begin{aligned}
E_\perp^\omega = j^\omega[\rho_{xx} - \rho_{zz}]sin(\theta)cos(\theta)+ \\
2j^\omega[cos^2(\theta) - sin^2(\theta)]B^2[T_{xzxz}sin(\alpha)cos(\alpha)cos^2(\beta) + T_{xzxy}sin(\alpha)sin(\beta)cos(\beta)] \\
+j^\omega sin(\theta)cos(\theta)B^2[T_{xxxx}sin^2(\alpha)cos^2(\beta) + T_{xxyy}sin^2(\beta) + T_{xxzz}cos^2(\alpha)cos^2(\beta)+ \\
T_{xxyz}cos(\alpha)sin(\beta)cos(\beta) - T_{zzxx}sin^2(\alpha)cos^2(\beta) - T_{zzyy}sin^2(\beta) - T_{zzzz}cos^2(\alpha)cos^2(\beta)]+ \\
R_{xzy}j^\omega[sin^2(\theta) + cos^2(\theta)]Bsin(\beta)
\end{aligned} \tag{S29}$$

Now, using: $E_\parallel^\omega = V_\parallel^\omega/L_\parallel$, $E_\perp^\omega = V_\perp^\omega/L_\perp$ and $j^\omega = I^\omega/A$, we can rewrite the expressions (S28) and (S29) in terms of the parameters that we directly measure in our experiments:

$$\begin{aligned}
V_\parallel^\omega = \frac{L_\parallel}{A}I^\omega[\rho_{xx}sin^2(\theta) + \rho_{zz}cos^2(\theta)]+ \\
2\frac{L_\parallel}{A}I^\omega sin(\theta)cos(\theta)B^2[T_{xzxz}sin(\alpha)cos(\alpha)cos^2(\beta) + T_{xzxy}sin(\alpha)sin(\beta)cos(\beta)]+ \\
\frac{L_\parallel}{A}I^\omega cos^2(\theta)B^2[T_{zzxx}sin^2(\alpha)cos^2(\beta) + T_{zzyy}sin^2(\beta) + T_{zzzz}cos^2(\alpha)cos^2(\beta)]+ \\
\frac{L_\parallel}{A}I^\omega sin^2(\theta)B^2[T_{xxxx}sin^2(\alpha)cos^2(\beta) + T_{xxyy}sin^2(\beta) + T_{xxzz}cos^2(\alpha)cos^2(\beta) + T_{xxyz}cos(\alpha)sin(\beta)cos(\beta)]
\end{aligned} \tag{S30}$$

$$\begin{aligned}
V_\perp^\omega = \frac{L_\perp}{A}I^\omega[\rho_{xx} - \rho_{zz}]sin(\theta)cos(\theta)+ \\
2\frac{L_\perp}{A}I^\omega[cos^2(\theta) - sin^2(\theta)]B^2[T_{xzxz}sin(\alpha)cos(\alpha)cos^2(\beta) + T_{xzxy}sin(\alpha)sin(\beta)cos(\beta)]+ \\
\frac{L_\perp}{A}I^\omega sin(\theta)cos(\theta)B^2[T_{xxxx}sin^2(\alpha)cos^2(\beta) + T_{xxyy}sin^2(\beta) + T_{xxzz}cos^2(\alpha)cos^2(\beta)+ \\
T_{xxyz}cos(\alpha)sin(\beta)cos(\beta) - T_{zzxx}sin^2(\alpha)cos^2(\beta) - T_{zzyy}sin^2(\beta) - T_{zzzz}cos^2(\alpha)sin^2(\beta)]+ \\
R_{xzy}\frac{L_\perp}{A}I^\omega[sin^2(\theta) + cos^2(\theta)]Bsin(\beta)
\end{aligned} \tag{S31}$$

To simplify the analysis, we will study the magnetoresistance and Hall contributions for the particular cases of current $j^\omega$ applied along $z$-axis ($\theta = 0°$):

$$V_\parallel^\omega = \frac{L_\parallel}{A} I^\omega \rho_{zz} + \frac{L_\parallel}{A} I^\omega B^2 [T_{zzxx} sin^2(\alpha) cos^2(\beta) + T_{zzyy} sin^2(\beta) + T_{zzzz} cos^2(\alpha) cos^2(\beta)] \quad (S32)$$

$$V_\perp^\omega = 2\frac{L_\perp}{A} I^\omega B^2 [T_{xzxz} sin(\alpha) cos(\alpha) cos^2(\beta) + T_{xzxy} sin(\alpha) sin(\beta) cos(\beta)] + R_{xzy} \frac{L_\perp}{A} I^\omega B sin(\beta) \quad (S33)$$

and along the $x$-axis ($\theta = 90°$):

$$V_\parallel^\omega = \frac{L_\parallel}{A} I^\omega \rho_{xx} + \frac{L_\parallel}{A} I^\omega B^2 [T_{xxxx} sin^2(\alpha) cos^2(\beta) + T_{xxyy} sin^2(\beta) +$$
$$T_{xxzz} cos^2(\alpha) cos^2(\beta) + T_{xxyz} cos(\alpha) sin(\beta) cos(\beta)] \quad (S34)$$

$$V_\perp^\omega = -2\frac{L_\perp}{A} I^\omega B^2 [T_{xzxz} sin(\alpha) cos(\alpha) cos^2(\beta) + T_{xzxy} sin(\alpha) sin(\beta) cos(\beta)] + R_{xzy} \frac{L_\perp}{A} I^\omega B sin(\beta) \quad (S35)$$

Moreover, to isolate the different components, we will consider 2 different rotating planes for the magnetic field: (1) $\alpha$-angle dependence with $\beta = 0°$, and (2) $\beta$-angle dependence with $\alpha = 0°$. Therefore, we have in total 4 cases:

(a) $\alpha$-angle dependence with $\beta = 0°$ and $\theta = 0°$:

$$V_\parallel^\omega = \frac{L_\parallel}{A} I^\omega \rho_{zz} + \frac{L_\parallel}{A} I^\omega B^2 [T_{zzxx} sin^2(\alpha) + T_{zzzz} cos^2(\alpha)] \quad (S36)$$

$$V_\perp^\omega = 2\frac{L_\perp}{A} I^\omega B^2 T_{xzxz} sin(\alpha) cos(\alpha) \quad (S37)$$

(b) $\beta$-angle dependence with $\alpha = 0°$ and $\theta = 0°$:

$$V_\parallel^\omega = \frac{L_\parallel}{A} I^\omega \rho_{zz} + \frac{L_\parallel}{A} I^\omega B^2 [T_{zzyy} sin^2(\beta) + T_{zzzz} cos^2(\beta)] \quad (S38)$$

$$V_\perp^\omega = R_{xzy} \frac{L_\perp}{A} I^\omega B sin(\beta) \quad (S39)$$

(c) $\alpha$-angle dependence with $\beta = 0°$ and $\theta = 90°$:

$$V_\parallel^\omega = \frac{L_\parallel}{A} I^\omega \rho_{xx} + \frac{L_\parallel}{A} I^\omega B^2 [T_{xxxx} sin^2(\alpha) + T_{xxzz} cos^2(\alpha)] \quad (S40)$$

$$V_\perp^\omega = -2\frac{L_\perp}{A} I^\omega B^2 T_{xzxz} sin(\alpha) cos(\alpha) \quad (S41)$$

(d) $\beta$-angle dependence with $\alpha = 0°$ and $\theta = 90°$:

$$V_\parallel^\omega = \frac{L_\parallel}{A} I^\omega \rho_{xx} + \frac{L_\parallel}{A} I^\omega B^2 [T_{xxyy} sin^2(\beta) + T_{xxzz} cos^2(\beta) + T_{xxyz} sin(\beta) cos(\beta)] \quad (S42)$$

$$V_\perp^\omega = R_{xzy} \frac{L_\perp}{A} I^\omega B sin(\beta) \quad (S43)$$

Therefore, from symmetry considerations, we have obtained the equations which describe the first-order electronic transport in Te.

**Experimental results**

We start exploring the zero-field resistivity components. For that purpose, we measured the longitudinal, $V_\parallel^\omega$ [Fig. S1(b)] and transverse, $V_\perp^\omega$ [Fig. S1(c)] first-harmonic voltage as a function of the $\theta$-angle, between the current $I^\omega$ and the chiral $z$-axis, in the absence of external magnetic fields. We represent the slope of each curve in front of the $\theta$-angle [Fig. 1(e)]. As observed, equations (S8) and (S9) perfectly fit the experimental data, demonstrating a precise control of the current directionality in our devices.

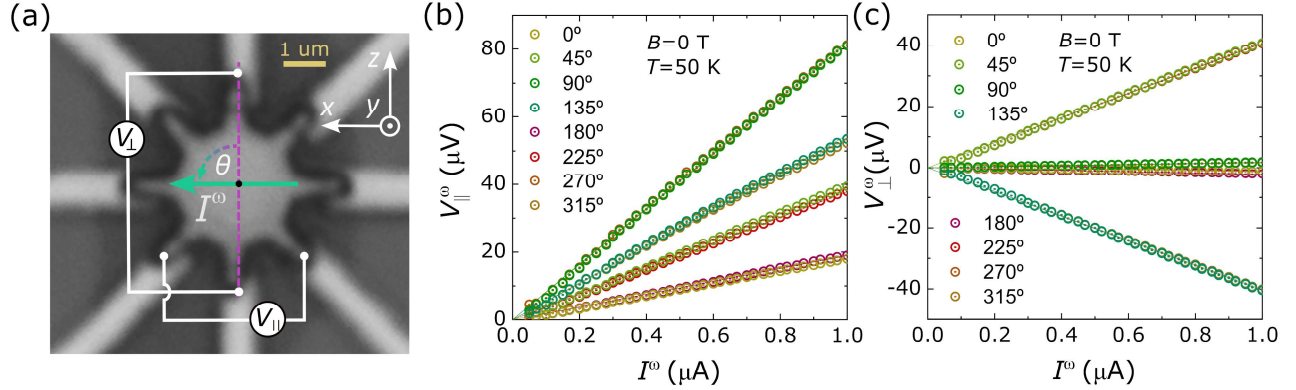

FIG. S1. Disc-like device and zero-field anisotropic resistivity. (a) Optical image of a typical device used in our experiments. The relative orientation between the current $I^\omega$ and the chiral axis is indicated. (b), (c) First-harmonic longitudinal $V_\parallel^\omega$ (b) and transverse $V_\perp^\omega$ (c) voltage as a function of $I^\omega$ at 50 K for different $\theta$ angles. The solid lines are linear fits to the experimental data.

Next, we measured the longitudinal magnetoresistance $MR = \frac{R_\parallel^\omega(B=9T) - R_\parallel^\omega(B=0T)}{R_\parallel^\omega(B=0T)}$ for both $\theta = 0°$ [Fig. 2(a)] and $\theta = 90°$ [Fig. 2(b)], as a function of (i) $\alpha$-angle with $\beta = 0°$ (blue curves) and (ii) $\beta$-angle with $\alpha = 0°$ (red curves). From equations (S36), (S38), (S40), and (S42) we can obtain directly the equations to fit our experimental data:

$$\theta = 0°, \beta = 0°, \alpha-\text{dep} \to MR = \frac{1}{\rho_{zz}} B^2 [T_{zzxx} sin^2(\alpha) + T_{zzzz} cos^2(\alpha)] \tag{S44}$$

$$\theta = 0°, \alpha = 0°, \beta-\text{dep} \to MR = \frac{1}{\rho_{zz}} B^2 [T_{zzyy} sin^2(\beta) + T_{zzzz} cos^2(\beta)] \tag{S45}$$

$$\theta = 90°, \beta = 0°, \alpha-\text{dep} \to MR = \frac{1}{\rho_{xx}} B^2 [T_{xxxx} sin^2(\alpha) + T_{xxzz} cos^2(\alpha)] \tag{S46}$$

$$\theta = 90°, \alpha = 0°, \beta-\text{dep} \to MR = \frac{1}{\rho_{xx}} B^2 [T_{xxyy} sin^2(\beta) + T_{xxzz} cos^2(\beta) + T_{xxyz} sin(\beta) cos(\beta)] \tag{S47}$$

As we can observe, these equations perfectly capture the experimental response [Fig. 2(a),(b)]. To further study the longitudinal magnetoresistance, we also measured the $MR$ as a function of the magnetic field when applied along $x$-, $y$-, and $z$-axis, both for $\theta = 0°$ [Fig. S2(a)] and $\theta = 90°$ [Fig. S2(b)]. Independently on the current direction, the material shows negative $MR$ when $B$ is aligned with the $z$-axis. The MR curves do not follow a parabolic dependence with $B$. However, they are even with respect to $B$, as expected. The later, differentiate a transverse signal arising from the magnetoresistance from a true Hall effect, which is odd with $B$ [46] (for more details, see Hall contribution section).

We also study the temperature dependence of the $MR$ between 10 K and 200 K. We focus on the $MR$ for $\theta = 0°$, when $B$ is applied along the $z$- as well as the $x$-axis. As we can observe in Fig. S2(c), when $B$ is applied along the $z$-axis, the MR magnitude decreases with increasing the temperature, but still negative between 0 and 9 T. In contrast, when $B$ is applied along the $x$-axis (Fig. S2(d)), we observe significant changes with temperature. At 10 K, it is negative until approximately 4 T and then becomes flat. At 50 K, it is negative until approximately 3 T and then becomes positive and parabolic. Finally, for 100 K and higher temperatures, the MR is positive and parabolic for the whole magnetic field range.

Now, we focus on the transverse response. For that purpose, we measured the transverse first-harmonic resistance $R_\perp^\omega = \frac{V_\perp^\omega}{I^\omega}$ for both $\theta = 0°$ [Fig. 2(c)] and $\theta = 90°$ [Fig. 2(d)], as a function of (i) $\alpha$-angle with $\beta = 0°$ (blue curves) and (ii) $\beta$-angle with $\alpha = 0°$ (red curves). From equations (S37), (S39), (S41), and (S43) we can obtain the equation to fit our experimental data:

$$\theta = 0°, \beta = 0°, \alpha-\text{dep} \to R_\perp^\omega = 2\frac{L_\perp}{A}B^2 T_{xzxz}\sin(\alpha)\cos(\alpha) \quad (S48)$$

$$\theta = 0°, \alpha = 0°, \beta-\text{dep} \to R_\perp^\omega = R_{xzy}\frac{L_\perp}{A}B\sin(\beta) \quad (S49)$$

$$\theta = 90°, \beta = 0°, \alpha-\text{dep} \to R_\perp^\omega = -2\frac{L_\perp}{A}B^2 T_{xzxz}\sin(\alpha)\cos(\alpha) \quad (S50)$$

$$\theta = 90°, \alpha = 0°, \beta-\text{dep} \to R_\perp^\omega = R_{xzy}\frac{L_\perp}{A}B\sin(\beta) \quad (S51)$$

We highlight that the component $T_{xzxz}$ describes what is usually called planar Hall effect. However, it is even with $B$ and it comes directly from the MR tensor, so it is not a true Hall effect. Moreover, the amplitude of the Planar Hall effect [Fig. 2(c),(d)] (blue curves) is much lower that the one of the longitudinal $MR$ [Fig. 2(a),(b)]. In the measurements of planar Hall effect in ferromagnets, the amplitude is linked to the amplitude of the longitudinal anisotropic $MR$, but here it is not, $T_{xzxz}$ is independent to $T_{xxxx}$, $T_{xxzz}$, $T_{zzxx}$, and $T_{zzzz}$. To further study the Hall component $R_{xzy}$, we also measured $R_\perp^\omega$ as a function of $B_y$ for $\theta = 0°$ [Fig. S2(e)] and $\theta = 90°$ [Fig. 2(f)]. From the positive sign, we identify that the transport is dominated by holes; and from the magnitude ($R_\perp^\omega = B_y/tn_p e$), we obtain the carrier density at 50 K: $n_p \approx 6.5 \times 10^{17}\text{cm}^{-3}$. Finally, we summarize in Table S1 the values of resistiviy, MR and Hall components obtained from the fittings of the experimental data.

TABLE. S1. Dimensions, zero-field resistivity, MR and Hall components at 50 K and 9 T.

| | | |
|---|---|---|
| Dimensions | $t$ ($\mu$m) | 0.11 |
| | $w$ ($\mu$m) | 0.79 |
| | $A$ ($\mu$m$^2$) | 0.087 |
| | $L_\parallel$ ($\mu$m) | 2.95 |
| | $L_\perp$ ($\mu$m) | 4.27 |
| Resistivity | $\rho_{xx}$ ($\Omega$m) | $1.81\ 10^{-3}$ |
| | $\rho_{zz}$ ($\Omega$m) | $4.09\ 10^{-4}$ |
| MR components | $T_{xxxx}$ (VmA$^{-1}$T$^{-2}$) | $(1.01 \pm 0.19)\ 10^{-7}$ |
| | $T_{xxyy}$ (VmA$^{-1}$T$^{-2}$) | $(4.94 \pm 0.22)\ 10^{-7}$ |
| | $T_{xxzz}$ (VmA$^{-1}$T$^{-2}$) | $(-1.178 \pm 0.021)\ 10^{-6}$ |
| | $T_{xxyz}$ (VmA$^{-1}$T$^{-2}$) | $\approx 0$ |
| | $T_{zzxx}$ (VmA$^{-1}$T$^{-2}$) | $(2.77 \pm 0.17)\ 10^{-8}$ |
| | $T_{zzyy}$ (VmA$^{-1}$T$^{-2}$) | $(1.360 \pm 0.022)\ 10^{-7}$ |
| | $T_{zzzz}$ (VmA$^{-1}$T$^{-2}$) | $(-1.658 \pm 0.017)\ 10^{-7}$ |
| | $T_{zzxx}$ (VmA$^{-1}$T$^{-2}$) | $(2.74 \pm 0.19)\ 10^{-9}$ |
| Hall component | $R_{xzy}$ (VmA$^{-1}$T$^{-1}$) | $(1.754 \pm 0.013)\ 10^{-6}$ |

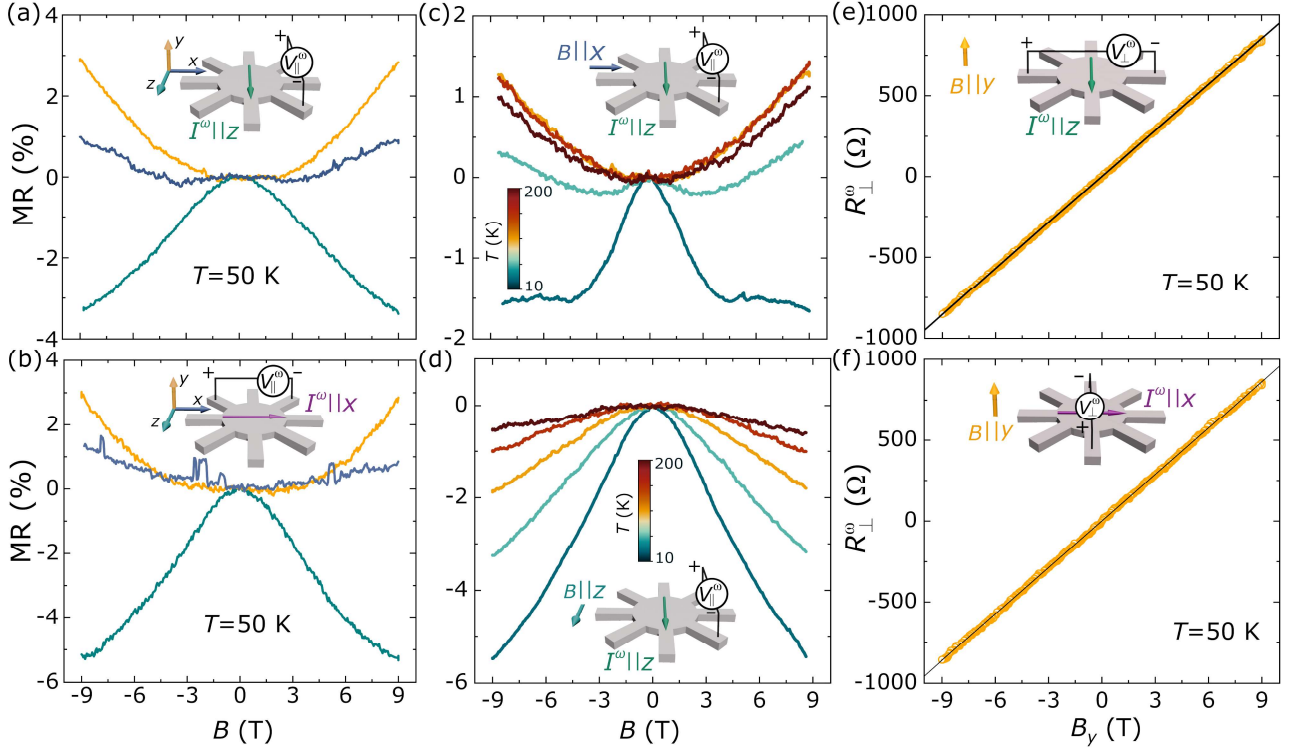

FIG. S2. Additional data of first-order electronic transport on Te. (a), (b) Longitudinal MR as a function of $B$ when applied along $x$-, $y$-, and $z$- axis, for (a) $\theta = 0°$ and (b) $\theta = 90°$ at 50 K. (c), (d), Temperature dependence of longitudinal MR for $\theta = 0°$ as a function of $B$ when applied along the (c) $x$-axis and (d) $z$-axis. (e), (f), $R_\perp^\omega$ as a function of $B_y$ for (e) $\theta - 0°$ and (f) $\theta - 90°$ at 50 K.

## Supplementary Section 3. Second-order electronic transport of Tellurium

The second-order nonlinear current density $j_i^{2\omega}$ in response to an electric field $E_{j,k}^\omega$ and magnetic field $B_l$ can be expressed through the material eMChA conductivity tensor $\sigma_{ijkl}$ as $j_i^{2\omega} = \sigma_{ijkl} E_j^\omega E_k^\omega B_l$. For Te with $P3_121$ or $P3_221$ point group symmetry, the eMChA tensor (considering the electric field and the current density in the $x-z$ plane and the magnetic field in all directions) has 8 independent non-zero elements [40]:

$$\sigma_{ijkl} = \begin{pmatrix} \sigma_{xxxx} & \sigma_{xzzx} & 0 \\ 0 & 0 & \sigma_{xxzy} \\ 0 & 0 & \sigma_{xxzz} \\ 0 & 0 & \sigma_{zxzx} \\ \sigma_{zxxy} & 0 & 0 \\ \sigma_{zxxz} & \sigma_{zzzz} & 0 \end{pmatrix} \tag{S52}$$

For an in-plane electric field $E_k^\omega = (E_x^\omega, 0, E_z^\omega)$, the second-order current density $j_i^{2\omega}$ is given by:

$$\begin{pmatrix} j_x^{2\omega} \\ j_z^{2\omega} \end{pmatrix} = \begin{pmatrix} \sigma_{xxxx} E_x^2 B_x + \sigma_{xzzx} E_z^2 B_x + 2\sigma_{xxzy} E_x E_z B_y + 2\sigma_{xxzz} E_x E_z B_y \\ 2\sigma_{zxzx} E_x E_z B_x + \sigma_{zxxy} E_x^2 B_y + \sigma_{zxxz} E_x^2 B_z + \sigma_{zzzz} E_x^2 B_z \end{pmatrix} \tag{S53}$$

In our experiments, we apply current and measure voltages. Therefore, we have to swap from conductivities to resistivities. Considering the electrical and eMChA conductivities, the current response is given by:

$$j_i = \sigma^0_{ij} E^\omega_j + \sigma_{ijkl} E^\omega_j E^\omega_k B_l \tag{S54}$$

We define the analog coefficients in resistivitiy as:

$$E_i = \rho_{ij} j^\omega_j + \rho_{ijkl} j^\omega_j j^\omega_k B_l \tag{S55}$$

Plugging Eq. (S55) back into Eq. (S54) and equating order by order, we can obtain the eMChA resistivity $\rho_{ijkl}$:

$$\sigma^0_{ij} \rho_{jklm} j^\omega_k j^\omega_l B_m + \sigma_{ijkl} \rho_{jm} j^\omega_m \rho_{kn} j^\omega_n B_l = 0 \tag{S56}$$

which gives after some relabeling:

$$\rho_{abcd} = -\sigma_{ijmd} \rho_{ia} \rho_{jb} \rho_{mc} \tag{S57}$$

Therefore, now we can write the second-order electric field $E^{2\omega}$ as a function of $\sigma_{ijkl}$:

$$\begin{pmatrix} E^{2\omega}_x \\ E^{2\omega}_z \end{pmatrix} = - \begin{pmatrix} \rho^3_{xx} \sigma_{xxxx} (j^\omega_x)^2 B_x + \rho_{xx} \rho^2_{zz} \sigma_{xzzx} (j^\omega_z)^2 B_x + 2\rho^2_{xx} \rho_{zz} \sigma_{xxzy} j^\omega_x j^\omega_z B_y + 2\rho^2_{xx} \rho_{zz} \sigma_{xxzz} j^\omega_x j^\omega_z B_y \\ 2\rho_{xx} \rho^2_{zz} \sigma_{zxzx} j^\omega_x j^\omega_z B_x + \rho^2_{xx} \rho_{zz} \sigma_{zxxy} (j^\omega_x)^2 B_y + \rho^2_{xx} \rho_{zz} \sigma_{zxxz} (j^\omega_x)^2 B_z + \rho^3_{zz} \sigma_{zzzz} (j^\omega_z)^2 B_z \end{pmatrix} \tag{S58}$$

For an in-plane current $j^\omega_i = j^\omega(sin(\theta), 0, cos(\theta))$ of amplitude $j^\omega$ and angle $\theta$ measured from the chiral axis ($z-$axis):

$$\frac{E^{2\omega}_x}{(j^\omega)^2} = -\rho^3_{xx} \sigma_{xxxx} sin^2(\theta) B_x - \rho_{xx} \rho^2_{zz} \sigma_{xzzx} cos^2(\theta) B_x$$
$$-2\rho^2_{xx} \rho_{zz} \sigma_{xxzy} sin(\theta) cos(\theta) B_y - 2\rho^2_{xx} \rho_{zz} \sigma_{xxzz} sin(\theta) cos(\theta) B_z \tag{S59}$$

$$\frac{E^{2\omega}_z}{(j^\omega)^2} = 2\rho_{xx} \rho^2_{zz} \sigma_{zxzx} sin(\theta) cos(\theta) B_x - \rho^2_{xx} \rho_{zz} \sigma_{zxxy} sin^2(\theta) B_y$$
$$-\rho^2_{xx} \rho_{zz} \sigma_{zxxz} sin^2(\theta) B_z - \rho^3_{zz} \sigma_{zzzz} cos^2(\theta) B_z \tag{S60}$$

Separating parallel and transverse components:

$$\frac{E^{2\omega}_\parallel}{(j^\omega)^2} = -\rho^3_{xx} \sigma_{xxxx} sin^3(\theta) B_x - \rho^3_{zz} \sigma_{zzzz} cos^3(\theta) B_z - [\rho^2_{xx} \rho_{zz} \sigma_{zxxy} + 2\rho^2_{xx} \rho_{zz} \sigma_{xxzy}] sin^2(\theta) cos(\theta) B_y$$
$$-[\rho^2_{xx} \rho_{zz} \sigma_{zxxz} + 2\rho^2_{xx} \rho_{zz} \sigma_{xxzz}] sin^2(\theta) cos(\theta) B_z - [\rho_{xx} \rho^2_{zz} \sigma_{xzzx} + 2\rho_{xx} \rho^2_{zz} \sigma_{zxzx}] sin(\theta) cos^2(\theta) B_x \tag{S61}$$

$$\frac{E^{2\omega}_\perp}{(j^\omega)^2} = \rho^2_{xx} \rho_{zz} \sigma_{zxxz} sin^3(\theta) B_z - \rho_{xx} \rho^2_{zz} \sigma_{xzzx} cos^3(\theta) B_x + [-\rho^3_{xx} \sigma_{xxxx} + 2\rho_{xx} \rho^2_{zz} \sigma_{zxzx}] sin^2(\theta) cos(\theta) B_x$$
$$+[\rho^3_{zz} \sigma_{zzzz} - 2\rho^2_{xx} \rho_{zz} \sigma_{xxzz}] sin(\theta) cos^2(\theta) B_z - 2\rho^2_{xx} \rho_{zz} \sigma_{xxzy} sin(\theta) cos^2(\theta) B_y + \rho^2_{xx} \rho_{zz} \sigma_{zxxy} sin^3(\theta) B_y \tag{S62}$$

For a rotating magnetic field: $\boldsymbol{B} = B(sin(\alpha)cos(\beta), sin(\beta), cos(\alpha)cos(\beta))$, we obtain:

$$\frac{E_\parallel^{2\omega}}{(j^\omega)^2} = -\rho_{xx}^3 \sigma_{xxxx} sin^3(\theta) B sin(\alpha) cos(\beta) - \rho_{zz}^3 \sigma_{zzzz} cos^3(\theta) B cos(\alpha) cos(\beta) - [\rho_{xx}^2 \rho_{zz} \sigma_{zxxy} +$$
$$2\rho_{xx}^2 \rho_{zz} \sigma_{xxzy}] sin^2(\theta) cos(\theta) B sin(\beta) - [\rho_{xx}^2 \rho_{zz} \sigma_{zxxz} + 2\rho_{xx}^2 \rho_{zz} \sigma_{xxzz}] sin^2(\theta) cos(\theta) B sin(\alpha) cos(\beta) -$$
$$[\rho_{xx} \rho_{zz}^2 \sigma_{xzzx} + 2\rho_{xx} \rho_{zz}^2 \sigma_{zxzx}] sin(\theta) cos^2(\theta) B sin(\alpha) cos(\beta) \tag{S63}$$

$$\frac{E_\perp^{2\omega}}{(j^\omega)^2} = \rho_{xx}^2 \rho_{zz} \sigma_{zxxz} sin^3(\theta) B cos(\alpha) cos(\beta) - \rho_{xx} \rho_{zz}^2 \sigma_{xzzx} cos^3(\theta) B sin(\alpha) cos(\beta) - [\rho_{xx}^3 \sigma_{xxxx} +$$
$$2\rho_{xx} \rho_{zz}^2 \sigma_{zxzx}] sin^2(\theta) cos(\theta) B sin(\alpha) cos(\beta) + [\rho_{zz}^3 \sigma_{zzzz} - 2\rho_{xx}^2 \rho_{zz} \sigma_{xxzz}] sin(\theta) cos^2(\theta) B cos(\alpha) cos(\beta) -$$
$$2\rho_{xx}^2 \rho_{zz} \sigma_{xxzy} sin(\theta) cos^2(\theta) B sin(\beta) + \rho_{xx}^2 \rho_{zz} \sigma_{zxxy} sin^3(\theta) B sin(\beta) \tag{S64}$$

As in previous sections, we study the particular cases of $j_i^\omega$ applied along x- ($\theta = 90°$) and z-axis ($\theta = 0°$). We obtain for $E_\parallel^{2\omega}$:

$$\theta = 0° \to \frac{E_\parallel^{2\omega}}{(j^\omega)^2} = -\rho_{zz}^3 \sigma_{zzzz} B cos(\alpha) cos(\beta) \tag{S65}$$

$$\theta = 90° \to \frac{E_\parallel^{2\omega}}{(j^\omega)^2} = -\rho_{xx}^3 \sigma_{xxxx} B sin(\alpha) cos(\beta) \tag{S66}$$

And for $E_\perp^{2\omega}$:

$$\theta = 0° \to \frac{E_\perp^{2\omega}}{(j^\omega)^2} = -\rho_{xx} \rho_{zz}^2 \sigma_{xzzx} B sin(\alpha) cos(\beta) \tag{S67}$$

$$\theta = 90° \to \frac{E_\perp^{2\omega}}{(j^\omega)^2} = -\rho_{xx}^2 \rho_{zz} \sigma_{zxxz} B cos(\alpha) cos(\beta) \tag{S68}$$

Finally, using: $E_\parallel^{2\omega} = V_\parallel^{2\omega}/L_\parallel$, $E_\perp^{2\omega} = V_\perp^{2\omega}/L_\perp$ and $j^\omega = I^\omega/A$, we can rewrite Eqs. (S63) and (S64) in terms of the parameters that we directly measure in our experiments. We obtain for $V_\parallel^{2\omega}$:

$$\theta = 0° \to \frac{V_\parallel^{2\omega}}{(I^\omega)^2} = -\frac{L_\parallel}{A^2} \rho_{zz}^3 \sigma_{zzzz} B cos(\alpha) cos(\beta) \tag{S69}$$

$$\theta = 90° \to \frac{V_\parallel^{2\omega}}{(I^\omega)^2} = -\frac{L_\parallel}{A^2} \rho_{xx}^3 \sigma_{xxxx} B sin(\alpha) cos(\beta) \tag{S70}$$

And for $V_\perp^{2\omega}$:

$$\theta = 0° \to \frac{V_\perp^{2\omega}}{(I^\omega)^2} = -\frac{L_\perp}{A^2} \rho_{xx} \rho_{zz}^2 \sigma_{xzzx} B sin(\alpha) cos(\beta) \tag{S71}$$

$$\theta = 90° \to \frac{V_\perp^{2\omega}}{(I^\omega)^2} = -\frac{L_\perp}{A^2} \rho_{xx}^2 \rho_{zz} \sigma_{zxxz} B cos(\alpha) cos(\beta) \tag{S72}$$

Which are equations (8), (9), (10) and (11) in the main text.

We investigate $V_\parallel^{2\omega}$ (Fig. 3a) and $V_\perp^{2\omega}$ (Fig. 3b) as a function of $\alpha$-angle at $\beta = 0°$ and $|B| = 9$ T, for current along z-axis ($\theta = 0°$) and x-axis ($\theta = 90°$). As observed, the equations obtained from our symmetry analysis perfectly capture the experimental response.

On the following, we will focus on the transverse eMChA, i.e. non-linear planar Hall effect, at $\theta = 90°$. In the main text, in order to further study this component, we recorded $V_\perp^{2\omega}$ for different currents at $B = 9$ T [Fig. 4(a)] and for different magnetic fields at $|I^\omega| = 1\mu A$ [Fig. 4(b)], as a function of $\beta$-angle at $\alpha = 0°$. Here, we also explore the current and field dependence of $V_\perp^{2\omega}$ but rotating $B$ in the $\alpha$-plane at $\beta = 0°$ [Fig. S3]. As observed, our equations perfectly fit the experimental data both in the $\alpha$- and $\beta$-angle dependence studies (solid lines in Fig. S3 and 4, respectively). We summarize in the main text the values of the eMChA components obtained from the fittings of the experimental data to equations (S69), (S70) and (S72).

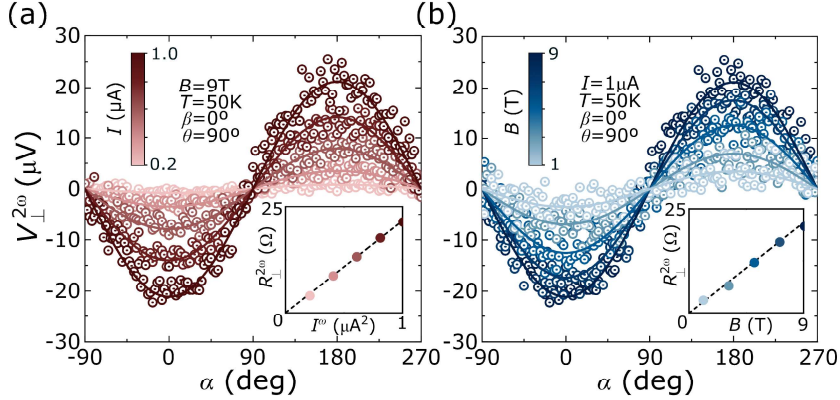

FIG. S3. Current and field dependence of the eMChA in Te. (a), (b) Second-harmonic transverse voltage ($V_\perp^{2\omega}$) when current is along the $x$-axis ($\theta = 90°$) and $\beta = 0°$ as a function of $\alpha$-angle for (a) different applied currents ($|I^\omega| = 0.2, 0.4, 0.6, 0.8,$ and $1.0$ $\mu A$) at 9 T and 50 K, (b) different external magnetic fields ($B = 1, 3, 5, 7, 9$ T) at 50 K and $|I^\omega| = 1\mu A$. Insets: (a) Current $I^\omega$, and (b) Magnetic field $B$ dependence of $R_\perp^{2\omega} = V_\perp^{2\omega}/I^\omega$. The values have been obtained from the fittings in the main panels (a) and (b) to Eq. (S72).

## Supplementary Section 4. How a combination of non-linear conductivity and ordinary Hall can give a contribution that mimics eMChA

If we also take into account the non-linear conductivity $\chi_{ijk}$ [16] and the Hall conductivity $\sigma_{ijk}^H$, we realize that more terms, by the combination of different contributions, appear into the resistivity. In general, the full current response is given by:

$$j_i = \sigma_{ij}^0 E_j^\omega + \chi_{ijk} E_j^\omega E_k^\omega + \sigma_{ijk}^H E_j^\omega B_k + \sigma_{ijkl} E_j^\omega E_k^\omega B_l \qquad (S73)$$

where we have considered the electrical conductivity $\sigma_{ij}^0$, the non-linear conductivity $\chi_{ijk}$ [16], the Hall conductivity $\sigma_{ijk}^H$, and the eMChA conductivity $\sigma_{ijkl}$. We define the analog coefficients in resistivitiy as:

$$E_i = \rho_{ij} j_j^\omega + \rho_{ijk}^{(2)} j_j^\omega j_k^\omega + R_{ijk} j_j^\omega B_k + \rho_{ijkl} j_j^\omega j_k^\omega B_l \qquad (S74)$$

Plugging Eq. (S74) back into Eq. (S73) and equating order by order, we can relate conductivities to resistivities. At zero-order in $B$ and first-order in $E$ we find $\sigma_{ij}^0 \rho_{jk} = \sigma_{ik}$. To second-order in $E$, we find:

$$\sigma_{ij}^0 \rho_{jkl}^{(2)} j_k^\omega j_l^\omega + \chi_{ijk} \rho_{jm} j_m^\omega \rho_{kn} j_n^\omega = 0 \qquad (S75)$$

which is equivalent to:
$$\rho^{(2)}_{jkl}j^\omega_k j^\omega_l = -\rho_{pi}\chi_{ijk}\rho_{jm}\rho_{kn}j^\omega_m j^\omega_n \tag{S76}$$

Now, relabeling indices:
$$\rho^{(2)}_{ijk} = -\rho_{il}\chi_{imn}\rho_{mj}\rho_{nk} \tag{S77}$$

To first-order in $B$ and $E$, we find:
$$\sigma^0_{ij}R_{jkl}j^\omega_k B_l + \sigma^H_{ijk}\rho_{jm}j^\omega_m B_k = 0 \tag{S78}$$

which gives:
$$R_{ijk} = -\rho_{il}\sigma^H_{lmk}\rho_{mj} \tag{S79}$$

Finally, to second-order in $E$ and first-order in $B$, we have:
$$\sigma^0_{ij}\rho_{jklm}j^\omega_k j^\omega_l B_m + \chi_{ijk}(\rho_{jl}j^\omega_l R_{kmn}j^\omega_m B_n + R_{jmn}j^\omega_m B_n \rho_{kl}j^\omega_l) + \sigma^H_{ijk}\rho^{(2)}_{jnm}j^\omega_n j^\omega_m B_k + \sigma_{ijkl}\rho_{jm}j^\omega_m \rho_{kn}j^\omega_n B_l = 0 \tag{S80}$$

which gives after some relabeling:
$$\rho_{abcd} = \left[\left(\chi_{ijk}\sigma^H_{lmd} + \chi_{imk}\sigma^H_{ljd} + \chi_{kjm}\sigma^H_{ild}\right)\rho_{kl} - \sigma_{ijmd}\right]\rho_{ia}\rho_{jb}\rho_{mc} \tag{S81}$$

To summarize, the relationship between resistivities and conductivities is given by:
$$\rho_{ij} = (\sigma^0_{ij})^{-1} \tag{S82}$$
$$\rho^{(2)}_{ijk} = -\rho_{il}\chi_{imn}\rho_{mj}\rho_{nk} \tag{S83}$$
$$R_{ijk} = -\rho_{il}\sigma^H_{lmk}\rho_{mj} \tag{S84}$$
$$\rho_{abcd} = \left[\left(\chi_{ijk}\sigma^H_{lmd} + \chi_{imk}\sigma^H_{ljd} + \chi_{kjm}\sigma^H_{ild}\right)\rho_{kl} - \sigma_{ijmd}\right]\rho_{ia}\rho_{jb}\rho_{mc} \tag{S85}$$

The last term in equation (S85) refers to the eMChA contribution, and the first three therms to the combination of non-linear conductivity and Hall effect. For Te, with $P3_121$ or $P3_221$ space group (twofold axis in the $x$-direction and chiral axis in $z$-direction), we have $\rho_{xx} = \rho_{yy}$ and $\rho_{zz}$ for resistivity, $\sigma^H_{zxy} = -\sigma^H_{zyx}$ and $\sigma^H_{xyz}$ for Hall, and $\chi_{xxx} = -\chi_{xyy} = -\chi_{yxy}$ and $\chi_{xyz} = -\chi_{yzx}$ for non-linear conductivity. Therefore, for our material we obtain:

$$\rho_{zzzz} = -\sigma_{zzzz}\rho^3_{zz} \tag{S86}$$
$$\rho_{xxxx} = -\sigma_{xxxx}\rho^3_{xx} \tag{S87}$$
$$\rho_{zxxz} = -\sigma_{zxxz}\rho_{zz}\rho^2_{xx} \tag{S88}$$
$$\rho_{zxxy} = -\sigma_{zxxy}\rho_{zz}\rho^2_{xx} + \chi_{xxx}\sigma^H_{zxy}\rho^3_{xx}\rho_{zz} \tag{S89}$$
$$\rho_{xzzx} = -\sigma_{xzzx}\rho_{xx}\rho^2_{zz} + 2\chi_{xzy}\sigma^H_{yzx}\rho_{xx}\rho^3_{zz} \tag{S90}$$

Taking into account all the contributions to $V^{2\omega}$:

$$\theta = 0° \rightarrow \frac{V^{2\omega}_\parallel}{(I^\omega)^2} = -\frac{L_\parallel}{A^2}\rho^3_{zz}\sigma_{zzzz}B\cos(\alpha)\cos(\beta) \tag{S91}$$

$$\theta = 90° \rightarrow \frac{V^{2\omega}_\parallel}{(I^\omega)^2} = -\frac{L_\parallel}{A^2}\rho^3_{xx}\sigma_{xxxx}B\sin(\alpha)\cos(\beta) - \frac{L_\parallel}{A^2}\rho^3_{xx}\chi_{xxx} \tag{S92}$$

$$\theta = 0° \rightarrow \frac{V^{2\omega}_\perp}{(I^\omega)^2} = -\frac{L_\perp}{A^2}\rho_{xx}\rho^2_{zz}\sigma_{xzzx}B\sin(\alpha)\cos(\beta) + 2\frac{L_\perp}{A^2}\chi_{xzy}\sigma^H_{yzx}\rho_{yy}\rho_{xx}\rho^2_{zz}B\sin(\alpha)\cos(\beta) \tag{S93}$$

$$\theta = 90° \rightarrow \frac{V^{2\omega}_\perp}{(I^\omega)^2} = -\frac{L_\perp}{A^2}\rho^2_{xx}\rho_{zz}\sigma_{zxxz}B\cos(\alpha)\cos(\beta) - \frac{L_\perp}{A^2}\rho^2_{xx}\rho_{zz}\sigma_{zxxy}B\sin(\beta)$$
$$+ \frac{L_\perp}{A^2}\chi_{xxx}\sigma^H_{zxy}\rho^3_{xx}\rho_{zz}B\sin(\beta) \tag{S94}$$

We note that the term $\frac{L_\parallel}{A^2}\rho_{xx}^3\chi_{xxx}$ in Eq. (S92) does not contribute to our magnetotransport experiments, because we measure at positive and negative magnetic fields and we calculate the half-difference. Therefore, terms at zero-order in $B$ are removed (for more details, see Supplementary Section 1, *III. Electrical measurements*).

## Supplementary Section 5.  Scaling law of eMChA

To unveil the microscopic mechanism behind a transport effect it is convenient to study the scaling law of the output voltage with the resistivity of the material. This methodology has been developed for many years and finally polished by Hou *et al.* [31] to study the anomalous Hall effect, and later adapted by Du *et al.* [5] to understand the non-linear Hall effect. Recently, Huang *et al.* proposed a scaling law for time-odd non-linear transport in magnetic materials [50]. In these systems, the role of internal magnetic vectors is similar to that of an external magnetic field in eMChA. Additionally, in eMChA, orbital magnetic field contributions should also be considered. Thus, by including this additional orbital contribution, the scaling law developed for time-odd non-linear transport in magnetic systems can be extended to apply to eMChA. The proposed scaling law for time-odd non-linear transport in magnetic systems is as follows:

$$\frac{V_\perp^{2\omega}}{(I^\omega)^2} = \gamma\rho_{xx}^{-1} + \delta + \xi\rho_{xx} + \zeta\rho_{xx}^2 + \eta\rho_{xx}^3 \tag{S95}$$

where $\eta$ is a resistivity-independent parameter, while $\gamma$, $\delta$, $\xi$ and $\zeta$ only depend on the residual resistivity of the material. The intrinsic Berry-connection polarizability [32] and the non-linear Drude term [33] are included in $\eta$ and $\xi$ parameters, respectively. The other contributions are a combination of skew-scattering and side-jump extrinsic mechanisms.

To incorporate the extra orbital magnetic field contributions in eMChA, we add the Berry curvature and Zeeman coupling mechanisms computed by Liu *et al.* [20]. The authors demonstrate that both terms scale quadratically with the scattering time and, therefore, must be included in the $\xi$ parameter. The Zeeman contribution is analogous to the non-linear Drude term in magnetic systems. Consequently, the scaling law for eMChA has the same form as the scaling law for time-odd non-linear transport in magnetic systems, with the intrinsic Berry-connection polarizability in the $\eta$ parameter and the Zeeman coupling and Berry curvature included now in the $\xi$ parameter.

To study the eMChA scaling law in our system, we modulate the resistivity by varying the sample temperature [Fig. 1(d)]. In the main text, we investigate the transverse component for $\theta = 90°$. Here, we also include the study of the longitudinal component for $\theta = 0°$. In this configuration, the scaling law can be written as:

$$\frac{V_\parallel^{2\omega}}{(I^\omega)^2} = \gamma\rho_{zz}^{-1} + \delta + \xi\rho_{zz} + \zeta\rho_{zz}^2 + \eta\rho_{zz}^3 \tag{S96}$$

We note that the Berry-connection polarizability mechanism was initially reported to be purely transverse [32]. However, later studies suggest that it also contributes to longitudinal non-linear transport [34, 35]. In our experiments, as we detail below, this contribution is not significant.

Hence, to unveil the dominant microscopic mechanism on Te, we study the temperature dependence of $V_\parallel^{2\omega}$ for $\theta = 0°$ as a function of the $\alpha$-angle at $B = 9$ T and $\beta = 0°$ [Fig. S4(a)]. Now, we can plot $V_\parallel^{2\omega}$ as a function of temperature [Fig. S4(b)], and as a function of the resistivity [Fig. S4(c)]. By fitting the experimental results to Eq. (S96), we recognize that the quadratic $\zeta$ term dominates (Table S2). Therefore, we can confirm that both longitudinal and tranverse eMChA in Te are dominated by extrinsic mechanisms. In table S2, we summarize the values of $\gamma$, $\delta$, $\xi$, $\zeta$ and $\eta$ obtained from the fittings of the experimental data to Eqs. (S95) and (S96).

TABLE. S2.  $\gamma$, $\delta$, $\xi$, $\zeta$ and $\eta$ values obtained from the fitting of the experimental data.

|  | $\gamma$ (mV$^2$A$^{-3}$) | $\delta$ (VA$^{-2}$) | $\xi$ (A$^{-1}$m$^{-1}$) | $\zeta$ ($m^{-2}V^{-1}$) | $\eta$ ($m^{-3}AV^{-2}$) |
| --- | --- | --- | --- | --- | --- |
| Transverse | $\approx 0$ | $(-9.6 \pm 2.9) \times 10^{-7}$ | $(2.89 \pm 0.46) \times 10^{-3}$ | $-2.44 \pm 0.17$ | $\approx 0$ |
| Longitudinal | $\approx 0$ | $(1.28 \pm 0.14) \times 10^{-5}$ | $(-8.84 \pm 0.78) \times 10^{-2}$ | $156 \pm 11$ | $\approx 0$ |

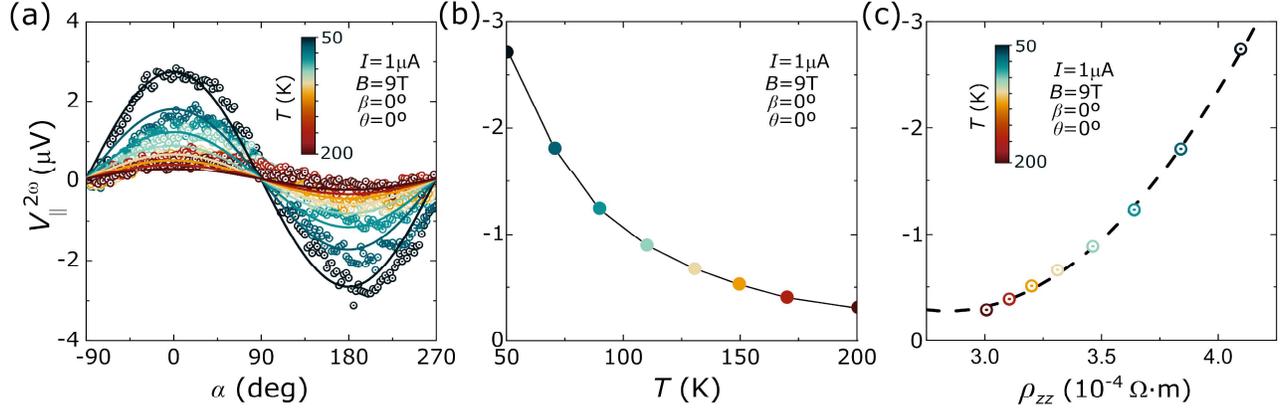

FIG. S4.  Scaling law of eMChA. Second-harmonic longitudinal voltage ($V_{\parallel}^{2\omega}$) when current is along the $z$-axis ($\theta = 0°$) as a function of (a) $\alpha$-angle for different temperatures ($T$ = 50, 70, 90, 110, 130, 150, 170 and 200 K), (b) temperature, and (c) resistivity along the z-axis ($\rho_{zz}$). The data in panel (b) is obtained from the fittings in panel (a). The measurements were performed at 9 T, $\beta = 0°$, and $|I^{\omega}| = 1$ $\mu$A.